# Trick or Treat? Free-ranging dogs use human behavioural cues for foraging

Rohan Sarkar[1], Sharmistha Maji[1], Tuhin Subhra Pal[1], Achal Dharmalal Rajratna[1], Avik Ghosh[2], Madhurima Roy[3,4], Sampurna Bag[5], Srijaya Nandi[1], Arpan Bhattacharyya[1], S. Sivasubramaniam[1], Avirup Chakraborty[6,7], Anindita Bhadra[1*]

[1] Department of Biological Sciences, Behaviour and Ecology Lab, Indian Institute of Science Education and Research Kolkata, Kolkata, West Bengal, India

[2] Department of Biological Sciences, Virology and Viral Immunology Lab, Indian Institute of Science Education and Research, Bhopal, Madhya Pradesh, India

[3] West Bengal Pollution Control Board, West Bengal, India

[4] Department of Environmental Science, Ballygunge Science College, University of Calcutta, Kolkata, West Bengal, India

[5] Department of Biological Sciences, Indian Institute of Science Education and Research, Berhampur, Odisha

[6] Adaptation Interface Lab, Ahmedabad University, Ahmedabad, Gujarat, India

[7] St. Xavier's College (Autonomous), Kolkata, India

**Corresponding author:**

*Anindita Bhadra

[1] Department of Biological Sciences, Behaviour and Ecology Lab, Indian Institute of Science Education and Research Kolkata, Kolkata, West Bengal, India, 741246

Email address: abhadra@iiserkol.ac.in

**Acknowledgement**

RS would like to thank the late, Mr. Anamitra Roy for his help with the fieldwork.

**Abstract**

Animals that display behavioural flexibility and adaptability thrive in urban environments, due to their ability to exploit novel anthropogenic resources. Since humans are an important component of such urban environments, animals that apply heterospecific learning in their decision-making are more likely to succeed as urban adapters. Free-ranging dogs, that have been living in human-dominated environments for centuries, are excellent urban adapters. In this study, we sought to understand the role and extent of human behavioural cues in decision-making during foraging by free-ranging dogs. We investigated whether these dogs were more attracted to items that humans appeared to be eating. When presented with a real and a fake biscuit, the dogs showed a clear preference for the food item. Between two identical biscuits, they chose the one that had been bitten by a human. However, when a fake biscuit was bitten and presented with a real one, the dogs failed to choose one over the other, suggesting a strong influence of the human-provided cue of biting over the natural cue of the smell of the food item. The dogs displayed left-bias during food choice across experimental conditions. These results demonstrate that dog foraging choices in urban environments are a mix of heterospecific learning and independent decision-making, highlighting an important facet behind their success in anthropogenic habitats. This also underscores the high level of dependence that free-ranging dogs have on humans in the urban habitat, not only as a source of food, but as an integral part of their ecological niche.

**Introduction**

Urban environments present formidable challenges to animals. In such environments, animals often come in contact with direct human presence or human generated products and processes. Urbanisation has transformed key ecological conditions and functions, especially food sources, dietary habits, and the process of acquiring food. Animals that thrive in such novel conditions display behavioural flexibility and traits that facilitate the use of anthropogenic resources, such as neophilia, boldness, and quick learning(Barrett et al., 2019). Although, urbanisation can have negative effects on some animals, it creates new opportunities for others. Human activities result in influx of nutrients in urban centres either intentionally (feeders) or unintentionally (garbage dumps, gardens)(Gaston et al., 2013). This, can in turn, lead to greater abundance of resources in terms of available prey. Furthermore, the easy accessibility and predictability of food resources in urban environments enables species with dietary flexibility and human tolerance to exploit these resources. Many mammalian carnivores are known to scavenge on human refuse and urban birds display more neophilia that results in successful exploitation of food resources in urban habitats(Bateman & Fleming, 2012; Tryjanowski et al., 2016).

This behavioural flexibility and adaptability displayed by urban animals, is in part, a result of social learning, wherein animals take cues from the behaviour of other animals to optimize their own(Ndousse et al., 2021). Such learning and innovative behaviour is seen at play when fish locate new food sources by observing the congregation of other members and house sparrows selecting their food by observing another member eating the same(Fryday & Greig-Smith, 1994; Laland, 2004). It has been hypothesized that the average fitness of a population comprised of selective social learners is higher than that of a population of individual learners. Animals are expected to use social learning more than individual learning to locate food and other resources when their habitat is changing at a moderate pace, a situation characteristic of urbanisation(Boyd R & Richerson, 2013; McKinney, 2008). That could be because individual exploration of environment is potentially costly and possibly dangerous and social learning allows animals to circumvent some of the costs(Dawson & Chittka, 2014; Lima & Dill, 1989). Furthermore, social learning has been shown to play an important role in developing diet choice and foraging strategies.

This learning by observing others can apply to both conspecifics and heterospecifics(Mason et al., 1984; May & Reboreda, 2005). In urban areas, animals come in frequent contact with humans and that may provide them with the opportunity to exercise heterospecific social learning. Herring gulls were found to use human handling as a cue to make choices regarding food(Goumas et al., 2020). Similar skills were found in other animals using different human cues(Schloegl et al., 2008; Smet & Byrne, 2013). The ability to understand, respond to, and learn from human cues has often been attributed to domestication and domestic animals often show the capacity to use human behavioural cues to make decisions and choices(Agnetta et al., 2000; Hare et al., 2002, 2010; Kaminski et al., 2005; Maros et al., 2008). But most of these studies have been done on captive, human-trained or human-reared animals with extensive experience with humans. But with increasing urbanisation, more and more wild and free-ranging animals are coming in contact with humans and anthropogenic resources and studying these animals in their natural habitat could help us investigate innate behavioural flexibility in animals and the effect of exposure to humans in ontogeny.

The free-ranging dog is an ideal model organism to test these effects as it occupies a unique niche. It is a group-living animal, living in close contact with humans(Sen Majumder et al., 2014). Thus, it can apply both conspecific and heterospecific social learning to increase feeding efficiencies and optimize diet choice. It is domesticated but has been living on its own for centuries, unrestricted in its movement or mating by humans which provides us with an opportunity to study the interplay of individual versus social learning influenced decision-making(Debroy, 2008; Serpell, 1995). They are opportunistic foragers, feeding from garbage dumps, dustbins and begging off of humans(Spotte, 2012). Thus, they primarily live on anthropogenic resources in highly variable, human influenced, urban environments. These characteristics and habitat conditions lead to their mixed relationship with humans(Butler & du Toit, 2002; Paul et al., 2016). Indeed, free-ranging dogs have shown remarkable socio-cognitive abilities in regards to humans as befits their multidimensional interactions. In fact, humans dominate the social interaction networks of urban free-ranging dogs(Bhattacharjee & Bhadra, 2020). They are able to process different human social communicative cues and modulate their behaviour accordingly(Bhattacharjee et al., 2018). They are highly sensitive to human attentional state as compared to pet and shelter dogs(Brubaker et al., 2019a). They are capable of following complex pointing cues from humans to locate hidden food rewards(Bhattacharjee et al., 2020). Interestingly, decision-making post the pointing gesture could be influenced by brief social petting. Dogs that received the petting followed even deceptive pointing cues whereas dogs that did not, discriminated between informative and deceptive pointing cues(Bhattacharjee & Bhadra, 2021). They can also learn to associate, recognize, and remember a person who provides them rewards over multiple interactions.(Nandi et al., 2024) Furthermore, free-ranging dogs in Chile were able to observe, follow, and learn from pedestrians that cross-walks were the safest areas to cross a road(Capell Miternique & Gaunet, 2020). Additionally, free-ranging dogs in Morocco were found to have been influenced by unfamiliar human actions in their foraging choices(Cimarelli et al., 2024).

In this between-subject study, we investigated whether observing human behavioural cues towards a particular item increased the probability of a free-ranging dog interacting with said item. This was measured by the number of times and the number of dogs approaching the item. By directly interacting with the object (biting), we might be communicating a strong signal of palatability to the dog that might in turn, influence their decision-making through local enhancement(Heyes, 1994). Visual lateralization has been known to influence foraging decisions and free-ranging dogs have shown a left-bias in a previous food preference task(Karenina et al., 2016; Sarkar et al., 2025; Vinassa et al., 2020). We hypothesized that a) free-ranging dogs would approach the bitten item more than the non-bitten item, regardless of whether the bitten item was a real food object or fake food object b) they would display a left-bias in approach as evidenced in previous studies in our lab, in absence of human cue, c) they would show quick decision-making and human cue/gesture would optimize such decisions d) gender of the dog has no effect whatsoever on their decision-making through social learning.

## Methods

### Study Areas

The study was conducted across 14 field sites in West Bengal. The field sites have been highlighted in the map (Figure 1). The study was carried out from August 2022 to March 2025.

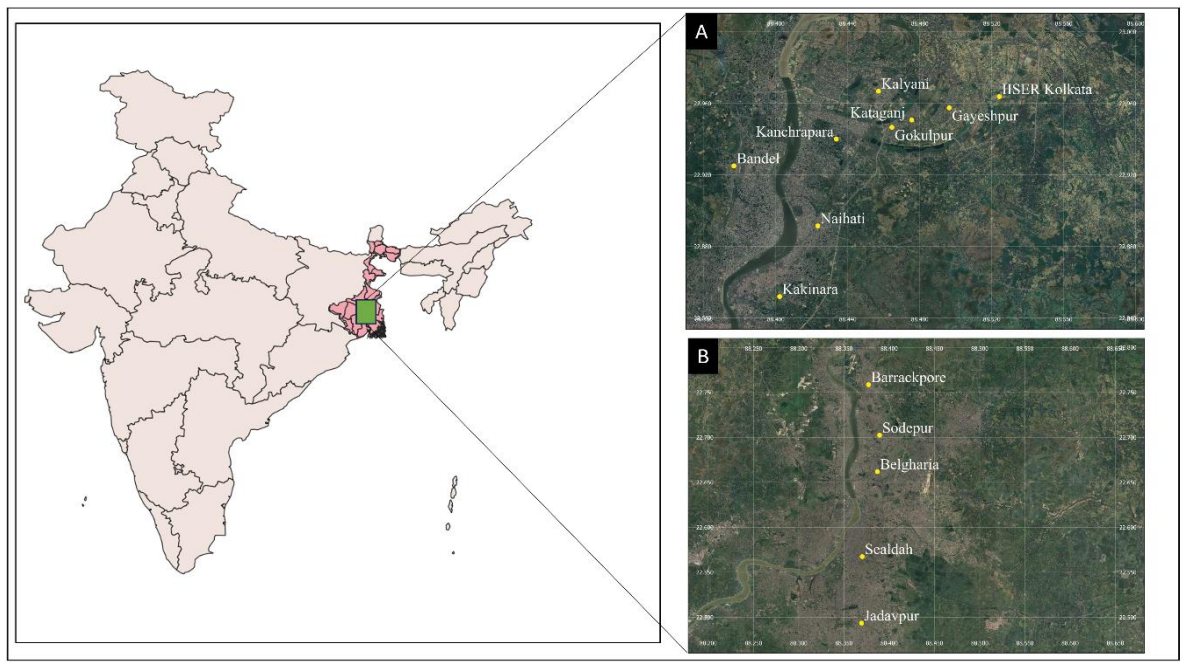

**Fig. 1** Map marking the field sites

### Subjects

We tested 300 dogs in total, out of which 258 dogs spread across the phases participated in the experiment and were included in the analysis. We only considered adult dogs for the experiment. Dogs were located randomly in different areas for the experiment. The experiment was carried out in different locations each day to avoid resampling.

**Phases**

There were four phases:

Bitten real: Both the biscuits provided to the dog were real. The experimenter bit off a tiny piece from one of the biscuits and then placed both of the biscuits on the ground. This was to check if dogs are attracted more to bitten than non-bitten food object.

Control real: Both the biscuits provided to the dog were real. No stimulus was provided. The biscuits were simply placed on the ground. This was to check the behaviour of the dog in the absence of a human gestural stimulus.

Bitten fake: One biscuit is real and the other is an identical but fake biscuit made of cardboard. The experimenter imitated the action of biting on the fake biscuit and then placed both of the biscuits on the ground. This was to check of dogs are attracted more the bitten fake food item than non-bitten food item and also to determine whether human behavioural cues alone could attract dogs towards any item, superseding natural cues like scent or if they are only attracted towards human behavioural cues when they are directed towards food items

Control fake: One biscuit is real and the other is fake. No stimulus was provided. The biscuits were simply placed on the ground. This was to check if the dogs showed any preference in the absence of a human gestural stimulus.

The type of biscuits provided are given in Figure 2.

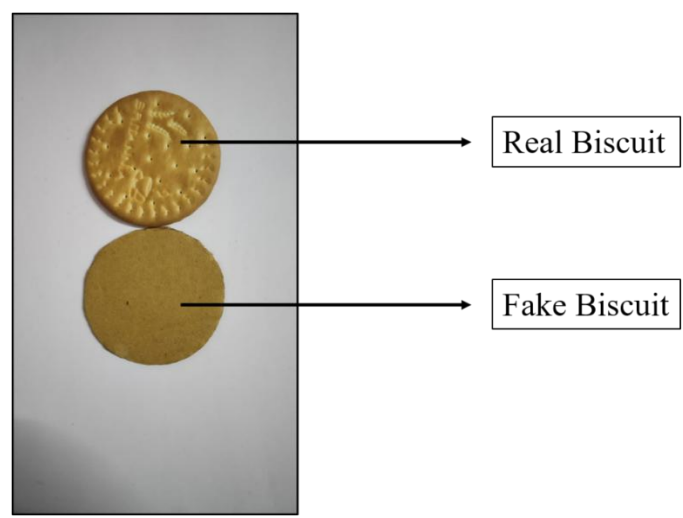

**Fig. 2** The two objects that were used as choices in the different phases

**Experimental Procedure**

Experimenters E1 and E2 would select a location and start walking. On encountering a dog, E1 would either bend or sit on their haunches to position themselves at the eye-level of the dog. First, they would display the two biscuits they held in their hand to the focal dog and then spread their arms apart to highlight the presence of choice. Depending on the phase, a stimulus would be provided. The two biscuits would then be placed on the ground at a distance of 1m from each other and 1-1.5m from the dog. The experimenter would then stand up, move 1-2 steps back and assume a neutral position. No other cues would be provided to the dog other than the focal stimulus. Each dog was subjected to three trials. In each of the trial a two-way choice test was provided. The dog was provided a maximum of one minute to make a choice. The experiment was stopped if no dog approached the set-up for one minute or made a choice, whichever occurred earlier. Choice was defined as dog sniffing the fake biscuit or eating the real biscuit.

**Behaviours and Analysis**

We considered two behaviours: sniffing and eating. A dog was said to be sniffing an object if it approached within 2-3cm of the object with the snout extended. If the dog picked up and put an object in its mouth, it was considered eating. Left and right sides of objects were considered from the point-of-view of the dogs.

Each phase had a different set of dogs. Thus, our study had a between-subjects design. All the analyses were done in R 4.3.1(R Core Team, 2025). Intercept-only regression models helped us compare frequencies against an expected ratio of chance for linked observations in case of multiple trials per individual dog. We carried out generalised linear mixed effects logistic regression for binomial data distribution using the lme4 packages for between group comparisons(Bates et al., 2015). We reported bias adjusted, model-estimated marginal means using the emmeans package (Russell V. Lenth, 2024; Searle et al., 1980) and log odds estimates. Random effect variables were included in the models, despite some models having low variance components to maintain fidelity between our models and the data-generating process. We carried out time-to-event analysis through mixed effects cox regression using the coxme package and compared latencies through Kaplan—Meier survival (event) estimates(Therneau T, 2023). Model diagnostics were carried out using the DHARMa package(Hartig, 2025). 20.79% of the data was re-coded by a second rater naïve to the experimental hypothesis. The inter-rater reliability was measured through the Cohen's kappa coefficient for the categorical variables and were found to be 0.846 – 0.983. For the quantitative data, we used intraclass correlation estimates based on a single rating, absolute agreement, two-way mixed effects model and the scores was 0.919.

**Results**

**Q1. Are dogs more attracted to the bitten than non-bitten food object?**

We ran a hierarchical, intercept-only, logistic regression to investigate the overall likelihood of choosing the bitten food object over the non-bitten food object across all three trials. The response variable, "approach" was binary (yes/no) and represented whether the dog approached, sniffed, and ate the bitten food object first. The random effects of "trial" was nested within "dog", nested within "place".

approach ~ 1 + (1 | place/dog/trial)

The model results showed that dogs preferred bitten food object over the non-bitten one across all three trials (0.67, p-value < 0.05). The model scored 0.86 on class separability (Figure 3).

**Q2. Are dogs more attracted to the bitten non-food object than non-bitten food object?**

We ran a hierarchical, intercept-only, logistic regression to investigate the overall likelihood of choosing the bitten non- food object over the non-bitten food object across all three trials. The response variable, "approach" was binary (yes/no) and represented whether the dog approached, sniffed, and interacted in any other way with the bitten non-food object first. The random effects of "trial" was nested within "dog", nested within "place".

approach ~ 1 + (1 | place/dog/trial)

The results of the regression showed that the approach and selection of an object, whether food or non-food was no different than that of a chance event (Figure 3).

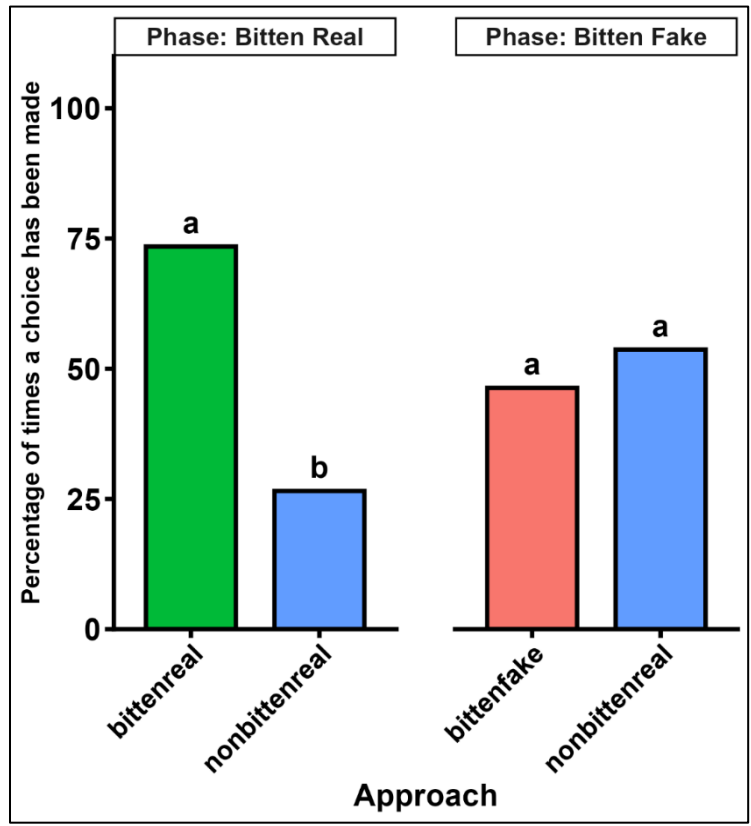

Fig. 3 Bar-graphs showing the percentage of times a particular choice was made across the two phases: Bitten Real and Bitten Fake

### Q3. Do dogs change their decision after choosing the fake biscuit?

We ran a hierarchical, intercept-only, logistic regression to investigate the overall likelihood that the dog would choose the non-food object once it has been deceived. The response variable, "changechoice" was binary (change/nochange) and represented whether after encountering the non-food object, the dog subsequently changed its choice in the next trial. The random effects of "trial" was nested within "dog", nested within "place".

changechoice ~ 1 + (1 | place/dog/trial)

The results of the regression showed that the decision of change or no change after encountering the non-food object was no different than that of a chance event.

### Q4. Do phase, gender and the side of the bitten object have an influence over the likelihood of approaching the bitten object?

In bitten real phase, of the 180 instances of the experiment being run, the bitten object was placed on the left and right sides on 89 and 91 instances respectively. Dogs, on the other hand, approached the left and right sides on 116 and 64 instances respectively. In the bitten fake phase, of the 177 instances of the experiment being run, the bitten object was placed on the left and right sides on 94 and 83 instances respectively. Dogs approached the left and right sides on 118 and 59 instances respectively. We ran a hierarchical logistic regression with a binary response variable, "approach" (yes/no), binary predictors, "phase", "gender" and "side", and random effect of trial nested within dog nested within place.

approach ~ 1 + phase + gender + side + (1 | place/dog/trial)

The results of the regression showed that dogs were more likely to approach the bitten object in the bitten real phase (0.88, p-value < 0.05) and when the object was on the left (1.35, p-value < 0.05). The model scored 0.78 on class separability. Gender had no influence on the likelihood of approach.

## Q5. How quickly does a dog make its choice?

We analysed whether there is a difference in latency of approach in dogs between different choices and gender. We defined latency as the time taken from when the objects were placed on the ground to the first sniff of one of the objects by the dog. We ran a nested, mixed effects cox proportional hazards model with random effects of "trial" nested within individual "dog" nested within "place". The maximum time provided to the dog to complete the experiment, 60s, was taken as the censored threshold value. The predictor variables were choice (bitten non-food object, bitten food object, non-bitten food object) and gender. The censored variable was sniff which was binary (1: dog sniffed, 0: dog did not sniff)

(latency, sniff) ~ choice + gender + (1|place/dog/trial)

The median latency for bitten non-food object, bitten food object, and non-bitten food object are 2.35, 1.50, 2.45 seconds respectively. The model estimated marginal means showed us that dogs that chose the bitten food object were likely to sniff earlier than those who chose non-bitten food object (non-bitten food object – bitten food object: -0.676, p-value < 0.05). There was no effect of one particular gender over another on latency.

## Q6. Are dogs more attracted to food object than non-food object without human gesture?

We ran a hierarchical, intercept-only, logistic regression to investigate the overall likelihood of choosing food object over the non-food object across all three trials without the presence of human gesture. The response variable, "approach" was binary (yes/no) and represented whether the dog approached and sniffed the food object first. The random effects of "trial" was nested within "dog", nested within "place". Subsequently, we ran another hierarchical, logistic regression to check the effect of side of the food-object (left/right) and gender on choice likelihood. The response variable and random effects were the same as the intercept-only model. The food object was kept 111 and 135 times on the left and right sides respectively.

approach ~ 1 + (1 | place/dog/trial)

approach ~ 1 + side + gender + (1 | place/dog/trial)

The intercept-only model results showed that dogs chose the food object over non-food object across 3 trials (0.67, p-value < 0.05; Figure 4). The model scored 1 on class separability. Placing the food object on the left increased its likelihood to be approached and sniffed at (1.79, p-value < 0.05). This model scored 0.91 on class separability. Gender had no effect.

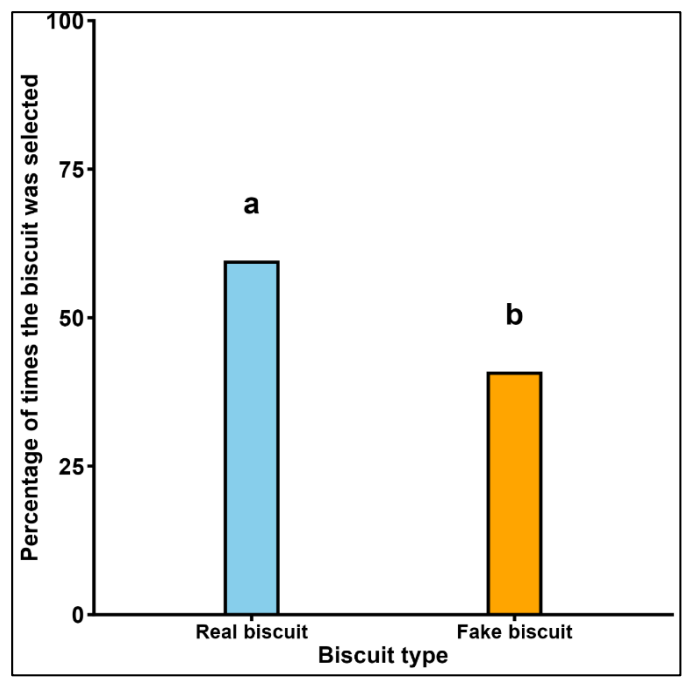

Fig. 4 A bar graph showing the percentage of times the real biscuit was chosen as opposed to the fake biscuit in the Control Fake phase

**Q7. Do dogs prefer one side over the other when both biscuits are real and non-bitten?**

Of the 152 instances, dogs went to the left and right side 90 and 62 times respectively. We ran a hierarchical, intercept-only, logistic regression to investigate the overall likelihood of choosing the food object on the left side versus the right across all three trials. The response variable, "firstapproachdir" was binary (left(1)/right(0)). The random effects of "trial" was nested within "dog", nested within "place".

firstapproachdir ~ 1 + (1 | place/dog/trial)

The model results showed that dogs chose the left biscuit significantly more than the right one across all three trials (0.44, p-value < 0.05; Figure 5). The model scored 0.95 on class separability.

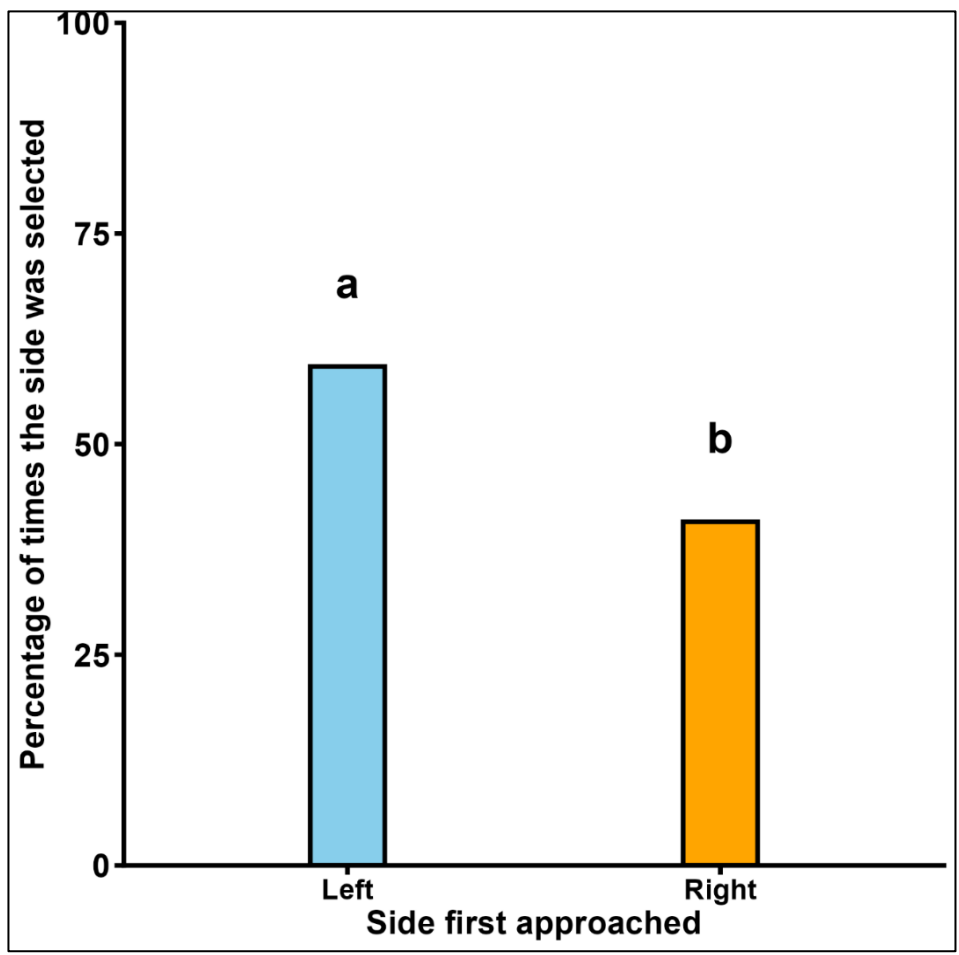

Fig. 5 A bar graph showing the percentage of times the dog approached one of the biscuits on either side in the Control Real phase

**Discussion**

The aim of the present study was to investigate whether free-ranging dogs used human behaviour to influence their foraging choices over and above food cues. By using a 2-choice test, we investigated whether human behavioural cues directed towards a food item would cause them to choose the same item. Dogs were more likely to approach and sniff at the food object that had been bitten by a human than an equally accessible, identical food that had not been bitten. This shows that dogs are attentive to human behavioural cue, similar to pet dogs(Brubaker et al., 2019b; Fugazza et al., 2016), and a human handled food is more attractive to them than a non-handled food.

To determine whether human behavioural cues alone could attract dogs towards any item, superseding natural cues like scent or if they are only attracted towards human behavioural cues when they are directed towards food items, we carried out another set of experiments using the same protocol but with the human behavioural cue being directed towards a non-food object (fake biscuit) which resembled the food item in shape and colour. A food object was the alternative choice. Dogs did not approach and sniff the non-food object above chance levels. On the other hand, without a human gesture to influence their decisions, dogs chose the food item significantly more than the non-food item. It is difficult to determine the reason behind the absence of a clear choice in the bitten fake phase, despite the dogs being able to discriminate between food and non-food object and being attentive to human behavioural cue. Dogs also did not show a clear strategy of maximising gains, by choosing food, after encountering the non-food object. Thus, there was no clear evidence of learning from previous experience. It can be hypothesized that a combination of the local and stimulus enhancement caused dogs to select either of the objects(Goumas et al., 2020; Hoppitt & Lala, 2013). These results share similarity with a study on the behaviour of companion dogs wherein half of them followed causally irrelevant human actions that did not lead to a food reward(Huber et al., 2018). Free-ranging dogs are neophilic and that may have also contributed to exploring the non-food object despite the presence of food(Bhattacharjee et al., 2024). Thus, our hypothesis about dogs following human behavioural cue to make their foraging decisions is partly validated.

In a social learning experiment in Morocco, free-ranging dogs were shown to preferably interact with a feeding box that had been previously interacted with by a human(Cimarelli et al., 2024). Our study, while working on similar principles is different in that the experimenter here directly interacted with the food and food-like objects. The current study supports the previous findings and further suggests that such heterospecific social learning behaviour is food-directed. In the absence of food, while human behavioural cue does have some effect on decision-making in dogs, it is not enough to supersede the food cues and that human cues may act as a supplemental factor in the presence of food cues.

This study is in line with earlier evidence that dogs pay attention to human demonstrators and can copy their action(Fugazza & Miklósi, 2015). The human demonstrator was unfamiliar to the dogs, showing a generalised reliance on humans for information. Using humans as a source of information is useful as the presence of feeders and casual feeding by humans make them a good predictor for availability of food in multiple places in India. Additionally, heterospecific learning confers multiple benefits as intraspecific learning but does not include competing for the same resources(Avarguès-Weber et al., 2013). Furthermore, free-ranging dogs have highlighted their ability to follow human gestures but also the behavioural flexibility to discriminate between informative and deceptive cues and take action accordingly should the gesture be deceptive(Bhattacharjee et al., 2017; Bhattacharjee & Bhadra, 2021).

Similar to a prior 2-choice experiment, we again found the evidence of laterality with dogs showing a left side bias(Sarkar et al., 2025). This further validates that dogs use relational spatial information while making foraging decisions in unpredictable environment. Dogs were also quick to approach the set-up giving further evidence to the speed-accuracy trade off strategy that they seem to follow in natural habitats.

In this study, we explored the influence of human behavioural cue on the decision-making of free-ranging dogs. Working with these animals present their own methodological challenges. With our set-up we were only able to test a subset of the population that was comfortable to the presence of a human to a certain degree. Animals that fled on human approach or were too anxious to approach could not be tested. We also could not control for prior association with humans and life experience. It is possible that dogs with positive experience with humans are more likely to follow their behavioural cues over independent decision-making, even if such cues lead to a wrong choice. Future research may consider prior human-dog interaction in the area as a factor for the decision-making. We also could not account for the hunger levels of the dogs. It maybe hypothesized that a well-fed or recently fed dog maybe less inclined towards foraging and more towards exploration. The current study adds to the growing research on heterospecific learning in free-ranging animals. It also demonstrates the interplay of heterospecific information gathering and independent decision making in free-ranging dogs. Understanding how human actions influence dog behaviour can be beneficial in raising awareness to reduce dog-human conflict and also aid in designing preventative measures.

**Data availability statement**

The datasets presented in this study can be found in online repositories. The names of the repository/repositories and accession number(s) can be found at: https://osf.io/uqyke/

**Author contributions**

RSa and ABha: project administration, and writing—reviewing and editing. RSa: conceptualization, formal analysis, methodology, visualization, and writing original draft. RSa, SSs: Data curation. RSa, SM, TsP, AdR, AG, MR, SB, SN, AB, AC: Investigation, ABha: funding acquisition, resources, supervision, and validation

**Ethics statement**

Ethical review and approval were not required for the animal study because the experiment did not involve any invasive procedure, and the food provided to the dogs were fit for human consumption. Dog feeding on streets is permitted by Prevention of Cruelty to Animals Act 1960 of the Parliament, and this experimental protocol did not need any additional clearance from the Institute ethics committee, as it did not violate the law.

**Conflict of interest**

The authors declare that the research was conducted in the absence of any commercial or financial relationships that could be construed as a potential conflict of interest.

**Funding**

RSa was supported by IISER Kolkata Institute fellowship. SM and TsP were supported by University Grants Commission India PhD fellowship. AG was supported by INSPIRE fellowship, Department of Science and Technology, India. SN was supported by the INSPIRE PhD fellowship, Department of Science and Technology, India. AB was supported by PhD fellowship under Council for Scientific and Industrial Research, India This study was supported by IISER Kolkata ARF and the Janaki Ammal National Women Bioscientist Award (BT/HRD/NBA-NWB/39/2020-21 (YC-1)) of the Department of Biotechnology, India.